\documentstyle[amsfonts,amstex,preprint,aps]{revtex}

\begin{document}
\title{Teleportation scheme implementing contextually the Universal Optimal Quantum
Cloning Machine and the Universal Not Gate. Complete experimental realization }
\author{M. Ricci,\ F. Sciarrino,\ C. Sias, and F. De Martini}
\address{Dipartimento di Fisica and \\
Istituto Nazionale per la Fisica della Materia\\
Universit\`{a} di Roma ''La Sapienza'', Roma, 00185 - Italy}
\maketitle

\begin{abstract}
By a significant modification of the standard protocol of quantum state
Teleportation two processes ''forbidden'' by quantum mechanics in their
exact form, the Universal NOT gate and the Universal Optimal Quantum Cloning
Machine, have been implemented {\it contextually} and {\it optimally} by a
fully {\it linear }method{\it .} In particular, the first experimental
demonstration of the \ {\it Tele-UNOT}\ {\it Gate}, a novel quantum
information protocol has been reported (cfr. quant-ph/0304070). A complete
experimental realization of the protocol is presented here.
\end{abstract}


Classical information is encoded in bits, viz. dichotomic variables that can
assume the values $0$ or $1.$ For such variables there are no theoretical
limitations as far as the ''cloning'' and/or ''spin-flipping'' processes are
concerned. However manipulations on the quantum analogue of a bit, a qubit,
have strong limitations due to fundamental requirements by quantum
mechanics. For instance, it has been shown that an arbitrary unknown qubit
cannot be perfectly cloned: $\left| \Psi \right\rangle \rightarrow \left|
\Psi \right\rangle \left| \Psi \right\rangle $, a consequence of the
so-called ``no cloning theorem'' \cite{1}. Another ''impossible'' device is
the quantum NOT gate, the transformation that maps any qubit into the
orthogonal one $\left| \Psi \right\rangle \rightarrow \left| \Psi ^{\perp
}\right\rangle $ \cite{2}. In the last years a great deal of theoretical
investigation has been devoted to finding the best approximation allowed by
quantum mechanics to these processes and to establish the corresponding
''optimal'' values of the ''fidelity'' $F<1$.\ This problem has been solved
in the general case \cite{3,4}. In particular, it was found that a
one-to-two universal optimal quantum cloning machine (UOQCM), i.e. able to
clone one qubit\ into two qubits ($1\rightarrow 2$), can be realized with a
fidelity $F_{CLON}=\frac{5}{6}$. The UOQCM has been experimentally realized
by a quantum optical ''amplification'' method, i.e. by associating the
cloning effect with a QED\ ''stimulated emission'' process \cite{5,6}. Very
recently it has been argued by \cite{6} that when the cloning process is
realized in a subspace $H$ of a larger nonseparable Hilbert space $H\otimes K
$ which is acted upon by a physical apparatus, the same apparatus performs 
{\it contextually} in the space $K$ the ''flipping'' of the input injected
qubit, then realizing a $(1\rightarrow 1)$ Universal-NOT gate (U-NOT) with a
fidelity $F_{NOT}=\frac{2}{3}$ \cite{4}. As an example, a UOQCM can be
realized on one output mode of a non-degenerate ''quantum-injected'' optical
parametric amplifier (QI-OPA), while the U-NOT transformation is realized on
the other mode \cite{7}.

In the present work this relevant, somewhat intriguing result is
investigated under a new perspective implying a {\it modified} Quantum State
Teleportation (QST)\ protocol, according to the following scheme. The QST\
protocol implies that an unknown input qubit $\left| \phi \right\rangle
_{S}=\alpha \left| 0\right\rangle _{S}+\beta \left| 1\right\rangle _{S}\ $is
destroyed at a sending place (Alice:\ ${\cal A}$) while its perfect replica
appears at a remote place (Bob:\ ${\cal B}$) via dual {\it quantum} and {\it %
classical} channels \cite{8}. Let us assume that Alice and Bob share the
entangled ''singlet''\ state $\left| \Psi ^{-}\right\rangle _{AB}=2^{-%
{\frac12}%
}\left( \left| 0\right\rangle _{A}\left| 1\right\rangle _{B}-\left|
1\right\rangle _{A}\left| 0\right\rangle _{B}\right) $, and that we want to
teleport the generic qubit $\left| \phi \right\rangle _{S}\equiv \left| \phi
\right\rangle $. The \ {\it singlet} state is adopted hereafter because its
well known invariance under SU(2) transformations will ensure the
''universality'' of the cloning and U-NOT processes, as we shall see \cite
{5,6,7}. The overall state of the system is then $\left| \Omega
\right\rangle _{SAB}$= $\left| \phi \right\rangle _{S}\left| \Psi
^{-}\right\rangle _{AB}$. Alice performs a Bell measurement by projecting
the joint state of the qubits $S$ and $A$ into the four Bell states $\left\{
\left| \Psi ^{-}\right\rangle _{SA},\left| \Psi ^{+}\right\rangle
_{SA},\left| \Phi ^{-}\right\rangle _{SA},\left| \Phi ^{+}\right\rangle
_{SA}\right\} $ spanning the 4-dimensional Hilbert space $H\equiv $ $%
H_{A}\otimes H_{S}$ and then sends the result to Bob by means of 2 bits of
classical information. In order to obtain $\left| \phi \right\rangle _{B}$,
Bob applies to the received state the appropriate unitary transformation $%
U_{B}$ according to the following protocol: $\left| \Psi ^{-}\right\rangle
_{SA}\rightarrow U_{B}={\Bbb I}$, $\left| \Psi ^{+}\right\rangle
_{SA}\rightarrow U_{B}=\sigma _{Z}$, $\left| \Phi ^{-}\right\rangle
_{SA}\rightarrow U_{B}=\sigma _{X}$, $\left| \Phi ^{+}\right\rangle
_{SA}\rightarrow U_{B}=\sigma _{Y}$ where the kets express the received
corresponding information,$\ I,\sigma _{Z},\sigma _{X}$ are respectively the
identity, phase flip, spin flip operators and $\sigma _{Y}=-i\sigma
_{Z}\sigma _{X}$. At last, the QST channel acts on the input state $\rho
_{S}\equiv \left| \phi \right\rangle \left\langle \phi \right| $ as the
identity operator: $E_{QST}(\rho _{S})={\Bbb \rho }_{B}$. In absence of the 
{\it classical} channel, i.e. of the appropriate $U_{B}$ transformation, the
apparatus realizes the map $E_{B}(\rho _{S})=%
{\frac12}%
{\Bbb I}_{B}$, corresponding to the {\it depolarization} channel $E_{DEP}$.
This is the worst possible case because any information about the initial
state $\rho _{S}$ is lost.

In order to implement the UOQCM and U-NOT at Alice's and Bob's sites, in the
present work we modify the QST\ protocol by performing a different \
measurement on the system $S$ $+$ $A$. This leads to a different content of
information to be transferred by the {\it classical} channel from ${\cal A}$
to ${\cal B}$. Precisely, the ''{\it Bell measurement}'', able to
discriminate between the 4 Bell states, is replaced here by a dichotomic ''%
{\it projective} {\it measurement}'' able to identify $\left| \Psi
^{-}\right\rangle _{SA}$ , i.e. the {\it anti-symmetric} subspace of $%
H\equiv $ $H_{A}\otimes H_{S}$, and its complementary {\it symmetric}
subspace. Let us analyze the outcomes of such strategy, schematically
represented by Figure 1. With a probability $p=\frac{1}{4}$ the $\left| \Psi
^{-}\right\rangle _{SA}$ is detected by ${\cal A}$. In this case the correct
QST channel $E_{QST}$ is realized. However, if this is not the case, with
probability $p=\frac{3}{4}$, Bob cannot apply any unitary transformation to
the set of the non identified Bell states $\left\{ \left| \Psi
^{+}\right\rangle _{SA},\left| \Phi ^{-}\right\rangle _{SA},\left| \Phi
^{+}\right\rangle _{SA}\right\} $, and then the QST channel implements the
statistical map $E\left( \rho \right) $ = $\frac{1}{3}\left[ \sigma _{Z}\rho
\sigma _{Z}+\sigma _{X}\rho \sigma _{X}+\sigma _{Y}\rho \sigma _{Y}\right] $%
. As we shall see, this map coincides with the map $E_{UNOT}\left( \rho
_{S}\right) $ which realizes the Universal Optimal-NOT gate, i.e. the one
that approximates {\it optimally} the flipping of one qubit $\left| \phi
\right\rangle $ into the orthogonal qubit $\left| \phi ^{\perp
}\right\rangle $, i.e. $\rho _{S}$ into $\rho _{S}^{\perp }\equiv \left|
\phi ^{\perp }\right\rangle \left\langle \phi ^{\perp }\right| $. Bob
identifies the two different maps realized at his site by reading the
information (1 bit) received by Alice on the classical channel. For example,
such {\it bit} can assume the value $0$ if Alice identifies the Bell state $%
\left| \Psi ^{-}\right\rangle _{SA}$ and $1$ if she does not. We name this
process $Tele-UNOT$ since it consists in the Teleportation of an {\it optimal%
} antiunitary map acting on any input qubit, the U-NOT Gate \cite{7,9}.

As already stated, it has been shown that in a bipartite entangled system
the {\it optimal} U-NOT Gate is generally realized {\it contextually}
together with an {\it optimal} quantum cloning process \cite{6,7}. Therefore
it is worth analyzing what happens when the overall state $\left| \Omega
\right\rangle _{SAB}$\ is projected onto the subspace orthogonal to $\left|
\Psi ^{-}\right\rangle _{SA}\left\langle \Psi ^{-}\right| _{SA}\otimes H_{B}$
by the projector: 
\begin{equation}
P_{SAB}=({\Bbb I}_{SA}-\left| \Psi ^{-}\right\rangle _{SA}\left\langle \Psi
^{-}\right| _{SA})\otimes {\Bbb I}_{B}.  \label{proiettore}
\end{equation}
This procedure generates the normalized state $\left| \widetilde{\Omega }%
\right\rangle \equiv P_{SAB}\left| \Omega \right\rangle _{SAB}=\sqrt{\frac{2%
}{3}}[\left| \xi _{1}\right\rangle _{SA}\otimes \left| 1\right\rangle
_{B}-\left| \xi _{0}\right\rangle _{SA}\otimes \left| 0\right\rangle _{B}]$
where: $\left| \xi _{1}\right\rangle _{SA}$= $\alpha \left| 0\right\rangle
_{S}\left| 0\right\rangle _{A}$ + $%
{\frac12}%
\beta (\left| 1\right\rangle _{S}\left| 0\right\rangle _{A}+\left|
0\right\rangle _{S}\left| 1\right\rangle _{A})$ and $\left| \xi
_{0}\right\rangle _{SA}$= $\beta \left| 1\right\rangle _{S}\left|
1\right\rangle _{A}$+ $%
{\frac12}%
\alpha (\left| 1\right\rangle _{S}\left| 0\right\rangle _{A}+\left|
0\right\rangle _{S}\left| 1\right\rangle _{A})$. By tracing this state over
the $SA$ and $B$ manifolds we get: $\rho _{SA}\equiv Tr_{B}\left| \widetilde{%
\Omega }\right\rangle \left\langle \widetilde{\Omega }\right| $ = $\frac{2}{3%
}\left| \phi \right\rangle \left| \phi \right\rangle _{SA}\left\langle \phi
\right| \left\langle \phi \right| _{SA}+\frac{1}{3}\left| \left\{ \phi ,\phi
^{\perp }\right\} \right\rangle _{SA}\left\langle \left\{ \phi ,\phi ^{\perp
}\right\} \right| _{SA}$ and $\rho _{B}^{out}\equiv Tr_{SA}\left| \widetilde{%
\Omega }\right\rangle \left\langle \widetilde{\Omega }\right| =\frac{1}{3}%
(2\rho _{S}^{\perp }+\rho _{S})$ , where: $\left| \left\{ \phi ,\phi ^{\perp
}\right\} \right\rangle _{SA}$= $2^{-%
{\frac12}%
}\left( \left| \phi ^{\perp }\right\rangle _{S}\left| \phi \right\rangle
_{A}+\left| \phi \right\rangle _{S}\left| \phi ^{\perp }\right\rangle
_{A}\right) $. We further project on the spaces $A$ and $B$ and obtain the
output states, which are found mutually equal: $\rho _{S}^{out}\equiv
Tr_{A}\rho _{SA}$ = $\frac{1}{6}(5\rho _{S}+\rho _{S}^{\perp })$ = $\rho
_{A}^{out}\equiv Tr_{S}\rho _{SA}$. At last, by these results the expected 
{\it optimal} values for the fidelities of the two ''forbidden''\ processes
are obtained: $F_{CLON}=Tr[\rho _{S}^{out}\rho _{S}]=Tr[\rho _{A}^{out}\rho
_{S}]=\frac{5}{6}$ and $F_{UNOT}=Tr[\rho _{B}^{out}\rho _{S}^{\perp }]=\frac{%
2}{3}\ $\cite{6,7}.

In the last years many experimental realizations of quantum state
Teleportation have been achieved \cite{10}. Recently the first ''active'',
i.e. {\it complete} version of the QST\ protocol was realized by our
Laboratory by physically implementing the Bob's unitary operation \cite{11}.
In the present experiment the input qubit was codified as the {\it %
polarization} state of a single photon belonging to the input mode $k_{S}$: $%
\left| \phi \right\rangle _{S}=\alpha \left| H\right\rangle _{S}+\beta
\left| V\right\rangle _{S}$, $\left| \alpha \right| ^{2}+\left| \beta
\right| ^{2}=1$: Figure 2. Here $\left| H\right\rangle $ and $\left|
V\right\rangle $ correspond to the horizontal and vertical polarizations,
respectively. In addition, an entangled pair of photons, $A$ and $B$ were
generated on the modes $k_{A}$ and $k_{B}$ by Spontaneous Parametric Down
Conversion (SPDC) in the {\it singlet} state: $\left| \Psi ^{-}\right\rangle
_{AB}$= $2^{-1/2}\left( \left| H\right\rangle _{A}\left| V\right\rangle
_{B}-\left| V\right\rangle _{A}\left| H\right\rangle _{B}\right) $. The {\it %
projective measurement} in the space $H=$ $H_{A}\otimes H_{S}$ was realized
by linear superposition of the modes $k_{S}$ and $k_{A}$ on a $50:50$ {\it %
beam-splitter, }$BS_{A}.$ Consider the overall output state realized on the
two output modes $k_{1}\,$and $k_{2}$ of $BS_{A}$ and expressed by a linear
superposition of the states: $\left\{ \left| \Psi ^{-}\right\rangle
_{SA},\left| \Psi ^{+}\right\rangle _{SA},\left| \Phi ^{-}\right\rangle
_{SA},\left| \Phi ^{+}\right\rangle _{SA}\right\} $. As it is well known,
the realization of the singlet $\left| \Psi _{SA}^{-}\right\rangle $ is
identified by the emission of \ one photon on each output mode of $BS_{A}$
while the realization of the set of the other three Bell states implies the
emission of 2 photons either on mode $k_{1}\,$or on mode $k_{2}$. The
realization of the last process, sometimes dubbed as ''Bose Mode
Coalescence'' (BMC) or ''Mode-occupation enhancement'' \cite{12} \ was
experimentally identified by the simultaneous clicking of the detectors $%
D_{2}$ and $D_{2}^{\ast }$ coupled to the output mode $k_{2}$ by the $50:50$
beam-splitter $BS_{2}$. The identical effect expected on mode $k_{1}$ was
not exploited, for simplicity. As shown by the theoretical analysis \ above,
the last condition implied the simultaneous realization in our experiment of
the U-NOT and UOQCM processes, here detected by a {\it post-selection}
technique. Interestingly enough, the {\it symmetry} of the projected
subspace of $H$ identified by BMC is implied by the intrinsic {\it Bose
symmetry} of the 2 photon Fock state realized at the output of $BS_{A}$.

The source of the SPDC process was a Ti:Sa mode-locked pulsed laser with
wavelength (wl) $\lambda =795nm$ and repetition rate $76MHz$. A weak beam,
deflected from the laser beam by mirror $M$, strongly attenuated by grey
filters $(At)$, delayed by $Z${\bf \ }$=2c\Delta t${\bf \ }via{\bf \ }a
micrometrically adjustable optical ''trombone'' was the source of \ the {\it %
quasi} single-photon state injected into $BS_{A}$ over the mode $k_{s}$. The
average number of injected photons was $\overline{n}$ $\simeq $ $0.1$ and
the probability of a spurious 2 photon injection was evaluated to be a
factor $0.05$ lower than for single photon injection. Different qubit states 
$\left| \phi \right\rangle _{S}$ were prepared via the optical Wave-Plate
(wp) $WP_{S}$, either a $\lambda /2$ or $\lambda /4$ wp. The main UV\ laser
beam with wl $\lambda _{p}=397.5nm$. generated by Second Harmonic
Generation, focused into a $1.5mm$ thick nonlinear (NL) crystal of $\beta $%
-barium borate (BBO) cut for Type II phase-matching, excited the SPDC\
source of the {\it singlet} $\left| \Psi ^{-}\right\rangle _{AB}$. The
photons $A$ and $B$ of each entangled pair were emitted over the modes $%
k_{A} $ and $k_{B}$ with equal wls $\lambda =795nm$. All adopted
photodetectors $(D)$ were equal SPCM-AQR14 single photon counters. One
interference filter with bandwidth $\Delta \lambda =3nm$ was placed in front
of each $D$ and determined the coherence time of \ the optical pulses: $\tau
_{coh}\simeq 350f\sec $.

In order to realize the $Tele-UNOT$ protocol, the BMC\ process on the output
mode $k_{2}$ of $BS_{A}$ was detected by a coincidence technique involving $%
D_{2}$ and $D_{2}^{\ast }$ [Fig.2, inset:\ (a)]. In order to enhance the ''%
{\it visibility}'' of the Ou-Mandel interference at the output of $BS_{A}$,
a $\pi -$preserving mode-selector $(MS)$ was inserted on mode $k_{2}$. On
the Bob's site, the polarization $\pi -state$ on the mode $k_{B}$ was
analyzed by the combination of the wp $WP_{B}$ and of the {\it polarization
beam splitter} $PBS_{B}$. For each input $\pi -state$ $\left| \phi
\right\rangle _{B}$, $WP_{B}$ was set in order to make the $PBS_{B}$ to
transmit $\left| \phi \right\rangle _{B}$ and to reflect $\left| \phi
^{\perp }\right\rangle _{B}$, by then exciting $D_{B}$ and $D_{B}^{\ast }$
correspondingly. First consider the Teleportation (QST)\ turned off, by
setting the optical delay $\left| Z\right| \gg c\tau _{coh\text{\ }}$i.e. by
spoiling the interference of photons $S$ and $A$ in $BS_{A}$. In this case,
since the states $\left| \phi \right\rangle _{B}$ and $\left| \phi ^{\perp
}\right\rangle _{B}$ were realized with the same probability on mode $k_{B}$%
, the rate of coincidences detected by the $D$-sets $[D_{B},D_{2},D_{2}^{%
\ast }]$ and $[D_{B}^{\ast },D_{2},D_{2}^{\ast }]\ $were expected to be
equal. By turning on the QST, i.e. by setting $\left| Z\right| <<c\tau
_{coh} $, the output state $\rho _{B}^{out}=(2\rho _{S}^{\perp }+\rho
_{S})/3 $ was realized implying a factor $R=2$ {\it enhancement} of the
counting rate $[D_{B}^{\ast },D_{2},D_{2}^{\ast }]$ and {\it no enhancement}
of $[D_{B},D_{2},D_{2}^{\ast }]$. The actual measurement of $R$ was carried
out,\ and the {\it universality} of the $Tele-UNOT$ process demonstrated, by
the experimental results shown in Figure 3. These 3-coincidence results,
involving the sets $[D_{B},D_{2},D_{2}^{\ast }]$ and $[D_{B}^{\ast
},D_{2},D_{2}^{\ast }]$ correspond to the injection of three different input
states: $\left| \phi \right\rangle _{S}=\left| H\right\rangle $, $\left|
\phi \right\rangle _{S}=2^{-1/2}(\left| H\right\rangle +\left|
V\right\rangle )$, $\left| \phi \right\rangle _{S}=2^{-1/2}(\left|
H\right\rangle +i\left| V\right\rangle )$. In Figure 3 the square and
triangular markers refer respectively to the $[D_{B}^{\ast
},D_{2},D_{2}^{\ast }]$ and $[D_{B},D_{2},D_{2}^{\ast }]$ coincidences
versus the delay $Z=2c\Delta t.$ We may check that the $Tele-UNOT$ process
only affects the $\left| \phi ^{\perp }\right\rangle _{B}$ component, as
expected. The {\it signal-to-noise} $(S/N)$\ ratio $R\;$was determined as
the ratio between the peak values, i.e. for $Z\simeq 0$, and the no BMC
enhancement values, i.e. for $\left| Z\right| \gg c\tau _{coh\text{\ }}$.
The experimental values of the $UNOT$ {\it fidelity\ }$F=R(R+1)^{-1}$ are: $%
F_{H}=0.641\pm 0.005$;$\ F_{H+V}=0.632\pm 0.006$; $F_{H+iV}=0.619\pm 0.006$,
for the three injection states $\left| \phi \right\rangle _{S}$. These
results, to be compared with the {\it optimal} value $F_{th}=2/3\approx
0.666 $ corresponding to the {\it optimal}$\;R=2$, have been evaluated by
taking into account the reduction, by a factor $\xi =0.7$, of the
coincidence rate due to the spurious simultaneous injection of two photons
on the mode $k_{S}$ and to the simultaneous emission of two SPDC\ pairs. The
factor $\xi $ was carefully evaluated by a side experiment involving the
detectors $D_{2}$and $D_{2}^{\ast }$. Note that the experimental Tele-UNot
peaks shown in Figure 3 indeed demonstrate the simultaneous, {\it contextual}
realization of the Quantum Cloning and UNot Processes on Alice' s and Bob's
sites, respectively.

For the sake of completeness, we wanted to gain insight in the linear {\it %
probabilistic} UOQCM\ process by investigating whether at the output of $%
BS_{A}$ the photons $S$ and $A$ were indeed left in the state $\rho
_{S}^{out}=(5\rho _{S}+\rho _{S}^{\perp })/6$ = $\rho _{A}^{out}$ after the
state projection. The cloning analysis was realized on the $BS_{A}$ output
mode $k_{2}$ by replacing the measurement set (a) with the (b) one, shown in
the inset of Figure 2. The polarization state on mode $k_{2}$ was analyzed
by the combination of the wp $WP_{C}$ and of the polarizer beam splitters $%
PBS$. For each input $\pi $-state $\left| \phi \right\rangle _{S}$, $WP_{C}$
was set in order to make $PBS^{\prime }$s to transmit $\left| \phi
\right\rangle _{S}$ and to reflect $\left| \phi ^{\perp }\right\rangle _{S}$%
. The ''cloned'' state $\left| \phi \phi \right\rangle _{S}$ was detected on
mode $k_{2}$ by a coincidence between the detectors $D_{C}$, $D_{C}^{\prime
} $. The generation of an entangled pair was assured by detecting one photon
on the mode $k_{B}$; in this case $PBS_{B}$ was removed and the field of
mode $k_{B}$ was coupled directly to $D_{B}$. Any coincidence detected by
the sets $[D_{C},D_{C}^{\prime },D_{B}]$ and $[D_{C},D_{C}^{\ast },D_{B}]$
implied the realization of\ the states $\left| \phi \phi \right\rangle _{S}$
and $\left| \phi \phi ^{\perp }\right\rangle _{S}$, respectively. In analogy
with the previous experiment, when $\left| Z\right| >>c\tau _{coh\text{\ }}$
the rate of coincidences detected by $[D_{C},D_{C}^{\prime },D_{B}]$ and $%
2\times \lbrack D_{C}^{\ast },D_{C}^{\prime },D_{B}]$ were expected to be
equal. By turning on the cloning machine, $\left| Z\right| <<c\tau _{coh}$,
an {\it enhancement }by a factor $R=2$ of the counting rate by $%
[D_{C},D_{C}^{\prime },D_{B}]$ and {\it no enhancement} by $%
[D_{C},D_{C}^{\ast },D_{B}]$ were expected. The experimental results of the $%
S/N$ ratio $R$, carried out by coincidence measurements involving $%
[D_{C},D_{C}^{\prime },D_{B}]$ and $[D_{C},D_{C}^{\ast },D_{B}]$ are
reported\ in the lower plots of Figure 3, again for the three different
input states: $\left| \phi \right\rangle _{S}=\left| H\right\rangle $, $%
\left| \phi \right\rangle _{S}=2^{-1/2}(\left| H\right\rangle +\left|
V\right\rangle )$, $\left| \phi \right\rangle _{S}=2^{-1/2}(\left|
H\right\rangle +i\left| V\right\rangle )$. The square and triangular markers
there refer respectively to the $[D_{C},D_{C}^{\prime },D_{B}]$ and $%
[D_{C},D_{C}^{\ast },D_{B}]$ coincidence plots Vs the delay $Z$. The
following values of the {\it cloning fidelity }$F=(2R+1)(2R+2)^{-1}$ were
found: $F_{H}=0.821\pm 0.003$;$\ F_{H+V}=0.813\pm 0.003$; $F_{H+iV}=0.812\pm
0.003$, to be compared with the {\it optimal} $F_{th}=5/6\approx 0.833$
corresponding to the limit $S/N$ value: $R=2$. As for the $Tele-UNOT$ gate
experiment, the factor $\xi =0.7$ has been corrected during the evaluation
of the fidelities. Finally note that, in the present context the entangled 
{\it singlet} state $\left| \Psi ^{-}\right\rangle _{AB}\ $was not strictly
necessary for the solely implementation of quantum cloning as we could model
the {\it local} effect of the {\it singlet} on the input mode $k_{A}$ by a 
{\it fully mixed }state $\rho _{A}=%
{\frac12}%
{\Bbb I}_{A}$ spanning a $2$ dimensional space. This has been realized
successfully in another experiment \cite{13}.

In summary, two relevant quantum information processes, forbidden by quantum
mechanics in their exact form are found to be connected {\it contextually}
by a modified quantum state teleportation scheme and can be ''optimally''
realized. \ At \ variance with previous experiments, the complete
implementation of the new protocol has been successfully performed by a
fully $linear$ optical setup. The results are found in full agreement with
theory.

This work has been supported by the FET European Network on Quantum
Information and Communication (Contract IST-2000-29681: ATESIT), by I.N.F.M.
(PRA\ ''CLON'')\ and by M.I.U.R.(COFIN 2002).

\vskip 8mm

\parindent=0pt

\parskip=3mm

Figure.1. (COLOR\ ONLINE) General scheme for the simultaneous realization of
the Tele-UNOT Gate and of the {\it probabilistic} Universal Quantum Cloning
Machine (UOQCM).

Figure.2. (COLOR\ ONLINE) Setup for the optical implementation of the {\it %
Tele-UNOT Gate} and the {\it probabilistic} UOQCM. The measurement setup
used for the verification of the cloning experiment is reported in the INSET
(b).

Figure.3.\ (COLOR\ ONLINE) Experimental results of the {\it Tele-UNOT Gate}
and the UOQCM for three input qubits. {\it Filled squares:} plots
corresponding to the ''correct'' polarization; {\it Open triangles}: plots
corresponding to the ''wrong'' polarization. The solid line represents the
best gaussian fit expressing the {\it correct} polarization.

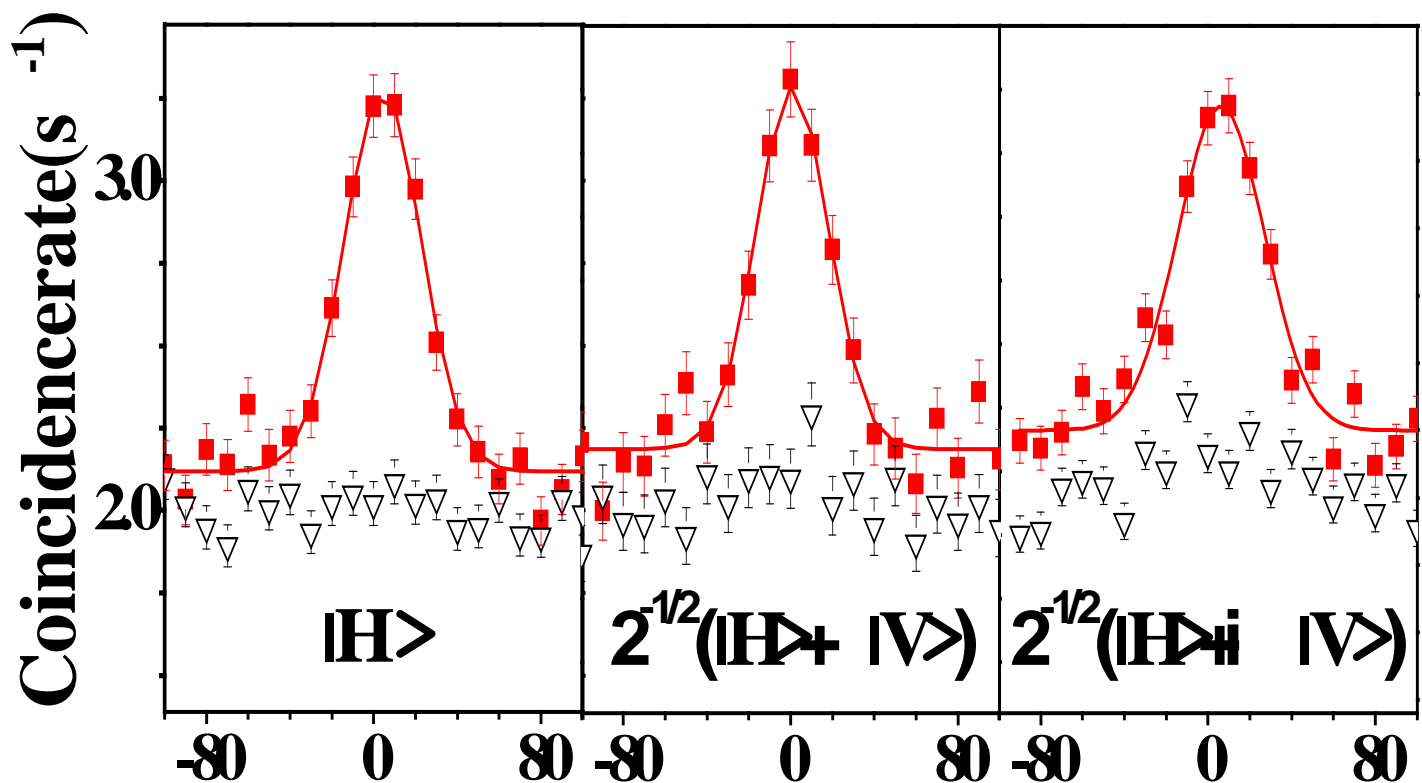
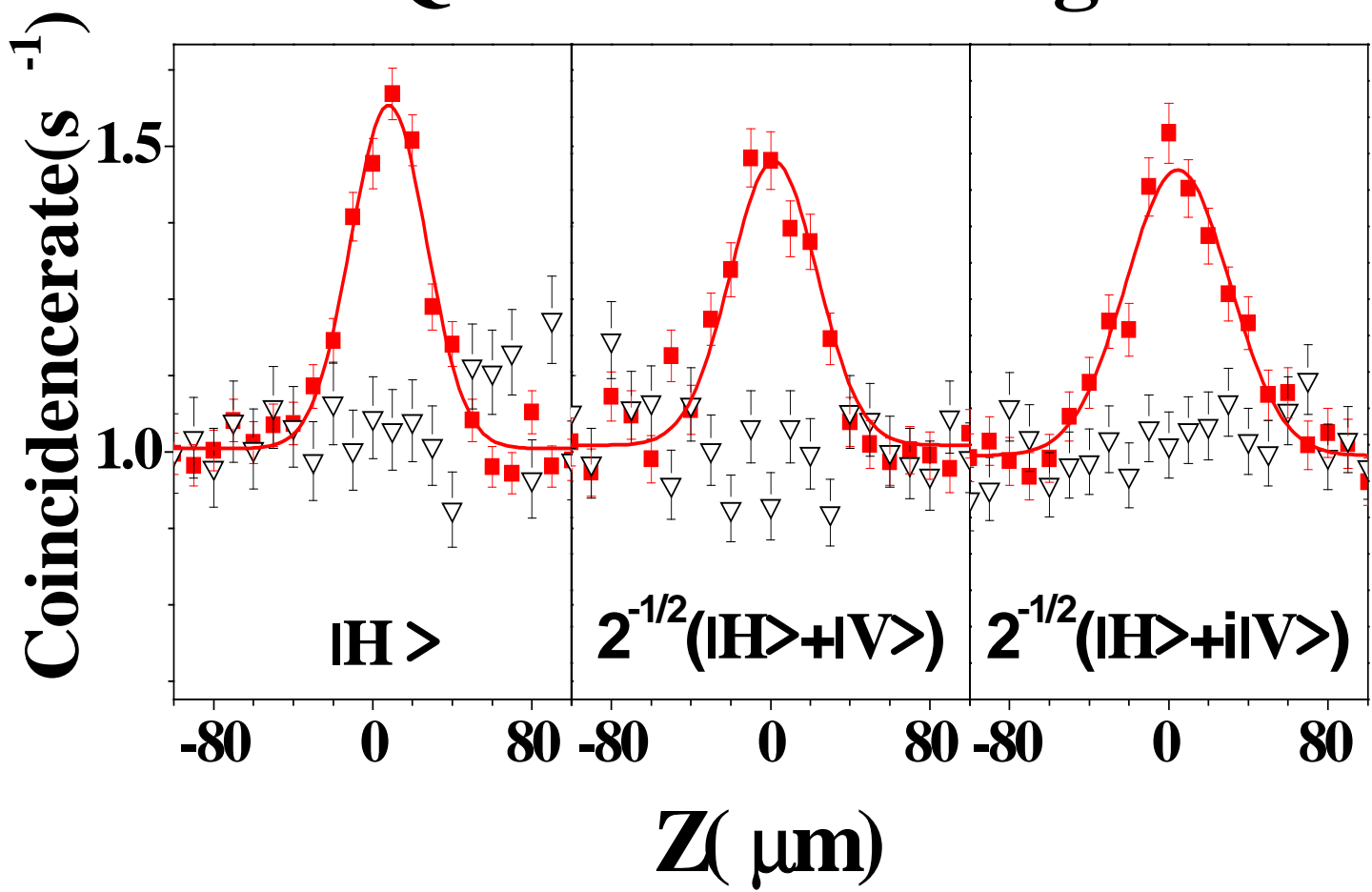
\end{document}